\documentclass{article}

\PassOptionsToPackage{numbers, compress}{natbib}

\usepackage{amsmath}
\usepackage{array}

    \usepackage[preprint]{neurips_2024}

\usepackage[utf8]{inputenc} 
\usepackage[T1]{fontenc}    
\usepackage{hyperref}       
\usepackage{url}            
\usepackage{booktabs}       
\usepackage{amsfonts}       
\usepackage{nicefrac}       
\usepackage{microtype}      
\usepackage{xcolor}         

\title{Can EEG resting state data benefit data-driven approaches for motor-imagery decoding?}

\author{
  Rishan Mehta \\
  Ahmedabad International School\\
  Ahmedabad, India \\
  \And
  Param Rajpura \\
  Human-AI Interaction (HAIx) Lab \\
  IIT Gandhinagar, India \\
  \AND
  Yogesh Kumar Meena \\
  Human-AI Interaction (HAIx) Lab \\
  IIT Gandhinagar, India \\
}

\begin{document}

\maketitle

\begin{abstract}

Resting-state EEG data in neuroscience research serve as reliable markers for user identification and reveal individual-specific traits. Despite this, the use of resting-state data in EEG classification models is limited. In this work, we propose a feature concatenation approach to enhance decoding models' generalization by integrating resting-state EEG, aiming to improve motor imagery BCI performance and develop a user-generalized model. Using feature concatenation, we combine the EEGNet model—a standard convolutional neural network for EEG signal classification with functional connectivity measures derived from resting-state EEG data. The findings suggest that although grounded in neuroscience with data-driven learning, the concatenation approach has limited benefits for generalizing models in within-user and across-user scenarios. While an improvement in mean accuracy for within-user scenarios is observed on two datasets, concatenation doesn't benefit across-user scenarios when compared with random data concatenation. The findings indicate the necessity of further investigation on the model interpretability and the effect of random data concatenation on model robustness.

\end{abstract}

\section{Introduction}

Data-driven approaches in Brain-Computer Interfaces (BCIs) are bringing neurotechnology closer to real-world applications\cite{schalk2024translation}. 
These interfaces have a broad range of applications, from restoring motor functions in individuals with severe motor impairments to providing new modes of interaction for able-bodied users in virtual environments.
The core functionality of BCIs relies on decoding neural activity, often recorded via electroencephalography (EEG), to infer user intent and translate it into actionable commands. 
Motor Imagery (MI), a paradigm in which users control external devices by imagining specific motor movements, such as moving a hand or foot is one of the most researched paradigm for BCIs with availability of large and multiple public datasets \cite{rajpura2024}. MI-BCIs leverage the ability of individuals to modulate their brain activity voluntarily, typically in the sensorimotor cortex, without executing actual movements. This allows for the control of assistive technologies through thought alone, making MI-BCIs particularly valuable for patients with conditions such as amyotrophic lateral sclerosis (ALS) or spinal cord injuries. However, the practical deployment of MI-BCIs is often challenged by the significant variability in EEG signals across sessions and subjects, driven by factors such as differences in electrode placement, neurophysiological conditions, and external noise.

Recent data-driven approaches with deep learning models trained on datasets from large number of participants show a promising approach for generalisability, requiring lesser or no calibration for new users \cite{zhang2021,kwon2020}.
The complex deep learning architectures at the same time are a hurdle to interpret the computations and decision making behind the model predictions \cite{rajpura2024}. 
This necessitates the incorporation of neurophysiological principles into model design and training to ensure that the models can generalize effectively beyond the specific conditions under which they were trained.

Recent efforts towards generalisation has focused on learning disentangled representations with complex encoder and decoder architectures \cite{jinpei2024,lies2022} to separate the signals from noise and user-specific signals. Another approach is towards the alignment of the covariance matrix that generalises the common spatial patterns for user-agnostic motor imagery decoding \cite{junqueira2024systematic,mellot2023}. 

Our research introduces an innovative approach to achieve generalization across sessions and users by utilizing resting-state EEG data. Prior studies have demonstrated the effectiveness of resting-state EEG as a distinctive biometric for individual identification \cite{Ma2015,choi2018individual,wang2019convolutional}, due to its reflection of spontaneous brain activity when no specific tasks are being performed. This brain activity captures the brain's inherent functional organization and has been associated with individual differences in cognitive and behavioral characteristics. Moreover, the correlated predictors of BCI performance, are also based on extracted features from resting state EEG data. \cite{tzdaka2020,trocellier2024validating,blankertz2010neurophysiological}. Therefore, integrating resting-state EEG features with task-related features from motor imagery (MI) may represent a potential pathway to enhance the generalization capabilities of decoding models. This approach is designed with a foundation in neurophysiology, requiring minimal modifications to the existing standard architecture.  

The key contributions of this paper are as follows: \begin{itemize} \item We introduce a novel approach that combines resting-state EEG features with task-related EEG features extracted from the decoding models across sessions and subjects in MI-BCI applications. \item We investigate the effectiveness of this approach and perform ablation studies with feature concatenation; to multiple datasets comprising multiple subjects and sessions. 
\end{itemize}


The rest of the paper is organized as follows. Section II details the dataset and preprocessing methods. Section III describes the model architecture and training procedures. Section IV presents the results and discusses the implications of our findings. Finally, Section V concludes with a summary of our contributions and future directions for research.

\section{Method}

\subsection{Dataset and Preprocessing}

The dataset used in this study consists of electroencephalogram (EEG) recordings from 87 individuals who participated in motor imagery (MI) tasks and resting-state conditions \cite{dreyer2023large}. The EEG data were collected using 27 electrodes placed with a 10-20 configuration system, with each electrode sampling at a rate of 512 Hz. The dataset consisted of 70 hours of recordings of 8-second long runs when participants performed motor imagery, i.e. imagining left and right-hand movements following a visual cue on the screen.

The dataset has two runs for each participant; "acquisition" runs were used for training and validation of the model, while the rest of the four runs are termed "online" runs. Following the benchmark set by Dreyer et al. \cite{dreyer2023large}, we use the two runs termed acquisition from each participant for training and the four online runs as test sets. A band-pass filter with a frequency range of 0.5-40 Hz was used to prepare the raw EEG data for analysis. Epochs, or time segments of EEG data, were created by segmenting the 3 seconds of data following the event marker at the onset of the visual cue for movement imagination. This standardized epoch length was maintained across all participants to ensure consistent temporal analysis. The resting state data was extracted from the first two seconds of the trial, where the participants focused on a fixation cue and were not explicitly instructed to rest. 

Another dataset, BNCI 2014 IIa Competition Dataset \cite{brunner2008bci}, consists of electroencephalogram (EEG) signals from 9 individuals who participated in motor imagery (MI) tasks and resting state conditions. The EEG data were collected using 22 electrodes, with each electrode sampling at a frequency of 250 Hz. The analysis involved two classes: right-hand and left-hand movement imagery, while feet and tongue movements were ignored. The paradigm used in this dataset also has a 3-second long trial. Therefore, the epochs were extracted using the same process and were pre-processed with a band-pass filter with the same range: 0.5-40 Hz. The resting state data was extracted from the first two seconds of the trial, where the participants focused on a fixation cue and were not explicitly instructed to rest. 

\subsection{Resting State Connectivity Analysis}

The preprocessing and analysis for the resting state data on both datasets were common. To analyze steady-state connectivity patterns from resting-state EEG data, we employed spectral connectivity measures, including coherence (COH) and phase-locking value (PLV) \cite{lachaux1999measuring}. These measures were computed across three primary frequency bands of interest: theta (4–8 Hz), alpha (8–13 Hz), and beta (13–30 Hz). The methodology consisted of several steps, including data preprocessing, frequency band definition, wavelet transformation, and spectral connectivity estimation.

The preprocessing phase involved using MNE-Python \cite{GramfortEtAl2013a} to process resting-state EEG data from each epoch spanning a time window from 0 to 2 seconds relative to the trial start onset. Baseline correction was applied over the entire epoch duration to normalize the data and minimize low-frequency drift or noise artefacts. A continuous wavelet transform (CWT) using Morlet wavelets \cite{tallon1997oscillatory} was then applied to decompose the EEG signals into these desired frequency bands. A set of center frequencies (freqs) was generated using a linear space between the minimum and maximum frequency limits, with 4 samples per Hz to ensure sufficient frequency resolution. The number of cycles for each frequency was set to half of the frequency value to optimize the balance between time and frequency localization. This approach enabled us to capture the oscillatory dynamics at multiple scales, which is crucial for characterizing the temporal structure of neural signals. Spectral connectivity was estimated using Coherence (COH) and phase-locking value (PLV) as connectivity metrics to evaluate both amplitude and phase coupling between different brain regions. 

COH is a measure of the linear relationship between two signals in the frequency domain, capturing both the amplitude and phase coupling across frequency bands.
\begin{equation}
        COH(f) = \frac{\left| E[S_{xy}(f)] \right|}{\sqrt{E[S_{xx}(f)] \cdot E[S_{yy}(f)]}}
\end{equation}
The cross-spectrum \( S_{xy}(f) \) is a measure of the spectral density of the correlation between two signals $x(t)$ and $y(t)$ at a specific frequency $f$. The auto-spectra $S_{xx}(f)$ and $S_{yy}(f)$ are the Fourier transforms of the autocorrelation functions of $x(t)$ and $y(t)$, respectively, and represent the power spectral densities of the signals.

Similarly, Phase-Locking Value (PLV) measures the consistency of the phase difference between two signals across multiple trials, independent of their amplitude. PLV ranges from 0 to 1, where 0 indicates no phase locking (random phase differences) and 1 indicates perfect phase synchronization (constant phase difference).
\begin{equation}
        PLV = \left| E \left[ \frac{S_{xy}(f)}{|S_{xy}(f)|} \right] \right|
\end{equation}

The resting state EEG analysis was performed on the segmented epochs, and the resulting connectivity matrices were averaged across each participant trial to obtain a representation of functional connectivity.

\subsection{Model Architecture and Training}

The EEGNet \cite{lawhern2018eegnet} model was employed to extract features from the EEG data using data-driven approach across multiple paradigms. EEGNet is a specialized neural network architecture designed to handle the unique characteristics of EEG signals. The model includes both temporal and spatial convolutional layers, which are optimized to capture relevant patterns from the multi-channel EEG data. Temporal convolutional layers focus on identifying patterns within the time domain of the signals, while spatial convolutional layers extract information based on the relationships between different EEG channels \cite{shukahara2021translation}. The EEGNet model was implemented using the Torcheeg framework \cite{zhang2024torcheeg}. During the training phase, the model parameters were optimized using the Adam optimizer. Key hyperparameters such as the learning rate and batch size were set to 0.0005 and 500 epochs, respectively. The model was trained to perform a binary classification task using a cross-entropy loss function. To ensure that the model would generalize well to new data, 5-fold cross-validation was performed.

\subsection{Feature Concatenation and Classification}
To concatenate the EEG resting state connectivity information, features obtained through cross-spectral analysis were integrated with the spatial-temporal features extracted by the EEGNet model after flattening the activations of the final convolution layer in the second block. This fused feature set was subsequently fed into a fully connected layer, followed by a non-linear activation and a final layer for linear classification. The modified EEGNet was then optimized to differentiate between various task conditions by utilizing the extensive information from the resting state connectivity and EEGNet-derived spatiotemporal features.

\subsection{Performance Evaluation}

The performance of the model was primarily evaluated using accuracy metrics, which measure the proportion of correctly classified epochs out of the total number of epochs. For the Dreyer et al. 2023 dataset, all reported performance is on the validation set from the "acquisition" runs, while "online" runs from all participants were used as the test set. The model was trained on the "acquisition" runs from all participants. Cross-validation accuracy is the accuracy on the left-out fold during training on "acquisition" runs. For the BCI IV IIa dataset, data is taken from all subjects and divided into five folds; 5-fold cross-validation performance is reported.

\section{Results}

\subsection{Impact of Feature Concatenation on Model Performance}

In our study, we investigated the effect of concatenating resting-state EEG features with task-related EEG features on the validation accuracy of the decoding models. The rationale behind this approach is that resting-state EEG captures the inherent neurophysiological characteristics of each subject, which are relatively stable over time and across different sessions. When these stable features are combined with task-related features that capture the specific neural responses associated with the MI tasks, the resulting feature set provides a more comprehensive representation of the neural activity.

This more comprehensive representation can lead to better model generalization. The model is trained on features that encapsulate the subject's stable neural traits and the dynamic neural responses to the tasks. As a result, the model can more accurately decode the intended motor imagery across different sessions and subjects, leading to higher validation accuracy.

Tables \ref{tab:comparison_mean_accuracy} and \ref{tab:INRIA_mean_accuracy} are the results for average training and validation accuracies for the two datasets: Dreyer et al. 2023 \cite{dreyer2023large} and BNCI 2014 IIa Competition Dataset \cite{brunner2008bci} respectively.
The Table \ref{tab:INRIA_mean_accuracy} provides an overview of the variation in validation accuracy before and after feature concatenation across four distinct experimental setups:

1) Dreyer et al. 2023 (Acquisition Data): This experiment focuses on the baseline data collected during the acquisition phase, which precedes the online experimental trials. This data was used to train the model and validated on five folds. The accuracy with standard deviation across five folds has been reported in the Table \ref{tab:INRIA_mean_accuracy}. 2) BCI Competition IV Dataset IIa: This row highlights the application of feature concatenation on the BCI IV IIa dataset, a widely recognized benchmark in brain-computer interface (BCI) research. The dataset, comprising multi-channel EEG recordings from multiple participants, was used to assess how concatenating additional features influenced model performance. 3) Dreyer et al. 2023 (Online Data): This experiment analyzes the data collected in real-time during the trials, known as the "online phase." Here, participants received real-time feedback based on their ongoing performance. The table reports the performance of the best model trained on acquisition data. 4) Dreyer et al. 2023 (LSO): This experiment employs the Leave-Subjects-Out (LSO) approach, where the model is trained on a randomly selected subset of participants and validated on the remaining participants. The LSO method ensures that the model's generalizability is tested across unseen participants. The performance of the model trained on 50 participants, evaluated on the last 9 participants with ID 50-59 is reported in Table \ref{tab:INRIA_mean_accuracy}.

\begin{table*}[h!]
\centering
\begin{tabular}{c|c|c|c}
\textbf{ID} & \multicolumn{1}{|p{3cm}}{\centering \textbf{Without Concatenation (\%)}}  & \multicolumn{1}{|p{3cm}}{\centering \textbf{With Concatenation (\%)}} & \multicolumn{1}{|p{3cm}}{\centering \textbf{Random Concatenation (\%)}} \\ \hline
Average Training Accuracy & 81.76 & 86.29 & 83.78 \\ 
k=1 & 77.46 & 77.46 & 75.14 \\ 
k=2 & 76.88 & 79.96 & 76.11 \\ 
k=3 & 74.52 & 77.41 & 79.73 \\ 
k=4 & 75.87 & 78.76 & 76.83 \\ 
k=5 & 76.64 & 78.76 & 77.41 \\ 
Mean Accuracy with SD & 76.27 $\pm$ 0.01135 & 78.47 $\pm$ 0.01066 & 77.05 $\pm$ 0.01723 \\
\end{tabular}
\caption{Average Training and Validation Accuracy on BCI Competition IV Dataset IIa}
\label{tab:comparison_mean_accuracy}
\end{table*}

\begin{table*}[h!]
\centering
\begin{tabular}{c|c|c|c}
\textbf{ID} & \multicolumn{1}{|p{3cm}}{\centering \textbf{Without Concatenation(\%)}}  & \multicolumn{1}{|p{3cm}}{\centering \textbf{With Concatenation (\%)}} & \multicolumn{1}{|p{3cm}}{\centering \textbf{Random Concatenation (\%)}} \\ \hline
Average Training Accuracy & 85.10 & 82.74 & 82.92 \\ 
k=1 & 84.23 & 82.74 & 77.08 \\ 
k=2 & 80.21 & 77.38 & 82.59 \\ 
k=3 & 73.36 & 81.55 & 82.74 \\ 
k=4 & 79.32 & 83.93 & 80.21 \\
k=5 & 78.87 & 82.74 & 81.25 \\
Mean Accuracy with SD & 79.20 $\pm$ 3.89 & 81.67 $\pm$ 2.54 & 80.77 $\pm$ 2.31 \\
\end{tabular}
\caption{Average Training and Validation Accuracy on Dreyer et al. 2023 dataset}
\label{tab:mean_accuracy}
\end{table*}


\begin{table*}[h!]
\centering
\begin{tabular}{c|c|c|c}
\textbf{ID} & \multicolumn{1}{|p{3cm}}{\centering \textbf{Without Concatenation(\%)}}  & \multicolumn{1}{|p{3cm}}{\centering \textbf{With Concatenation (\%)}} & \multicolumn{1}{|p{3cm}}{\centering \textbf{Random Concatenation (\%)}} \\ \hline
Dreyer et al. 2023(Acq) & 79.20 $\pm$ 03.89 & 81.67 $\pm$ 02.54 & 80.77 $\pm$ 02.31 \\ 
BCI IV IIa & 76.27 $\pm$ 01.13 & 78.47 $\pm$ 01.07 & 77.05 $\pm$ 01.72 \\ 
Dreyer et al. 2023(Online) & 85.10 & 82.74 & 82.92 \\ 
Dreyer et al. 2023(LSO) & 74.04 & 72.57 & 73.67 \\
\end{tabular}
\caption{Mean accuracy with standard deviation across five iterations of training EEGNet with different configurations on Dreyer et al. 2023 Dataset \cite{dreyer2023large} and BCI IV IIa Competition Dataset}
\label{tab:INRIA_mean_accuracy}
\end{table*}



\section{Discussion and Conclusion}



\subsection{Interpreting the results and its impact}


Even though we attempted our approach grounded in neuroscience, the results show that it is not beneficial to concatenate resting state data. Even though the validation accuracy while concatenating resting state data was higher than without concatenation, the difference is negligible considering the standard deviation. The concatenation approach uses an additional non linear activation layer after combining the resting state data with activity data. The increase in accuracy while using random concatenation can be attributed to the patterns in random data and the fully connected layer with non-activation layer.




\subsection{Potential next steps}
While the current work demonstrates the ineffectiveness of the approach using resting state data concatenation, several areas for improvement and future exploration have been identified. One key limitation lies in not optimally targeting the specific layer within existing architectures, such as EEGNet, for concatenation of the resting-state data. A more refined approach could yield better benefits, which involves retaining generic features from earlier layers and strategically concatenating resting-state-specific information at later layers. Further, defining a meta-learning architecture that systematically incorporates resting-state EEG information could enhance model adaptability across different users and conditions.

Additionally, exploring conditional variational autoencoder (CVAE) architectures presents a promising direction for future research. CVAEs could offer a generative-discriminative framework for user-independent EEG decoding, leveraging resting-state data to improve generalizability across subjects.

\subsection{Conclusion}
Overall, this study underscores the investigation and the effectiveness of combining EEG resting-state data with advanced data-driven models to enhance MI classification performance. The findings suggest that feature concatenation is not a promising approach for generalisation of models. However, more complex approaches could leverage resting state information to impact the field of brain-computer interfaces by providing more personalized and effective solutions for real-world applications.

\bibliographystyle{abbrv}
\bibliography{restEEGconcat}

\begin{thebibliography}{10}

\bibitem{blankertz2010neurophysiological}
B.~Blankertz, C.~Sannelli, S.~Halder, E.~M. Hammer, A.~K{\"u}bler, K.-R. M{\"u}ller, G.~Curio, and T.~Dickhaus.
\newblock Neurophysiological predictor of smr-based bci performance.
\newblock {\em Neuroimage}, 51(4):1303--1309, 2010.

\bibitem{lies2022}
L.~Bollens, T.~Francart, and H.~V. Hamme.
\newblock Learning subject-invariant representations from speech-evoked eeg using variational autoencoders.
\newblock In {\em ICASSP 2022 - 2022 IEEE International Conference on Acoustics, Speech and Signal Processing (ICASSP)}, pages 1256--1260, 2022.

\bibitem{brunner2008bci}
C.~Brunner, R.~Leeb, G.~M{\"u}ller-Putz, A.~Schl{\"o}gl, and G.~Pfurtscheller.
\newblock Bci competition 2008--graz data set a.
\newblock {\em Institute for knowledge discovery (laboratory of brain-computer interfaces), Graz University of Technology}, 16:1--6, 2008.

\bibitem{choi2018individual}
G.-Y. Choi, S.-I. Choi, and H.-J. Hwang.
\newblock Individual identification based on resting-state eeg.
\newblock In {\em 2018 6th International conference on brain-computer interface (BCI)}, pages 1--4. IEEE, 2018.

\bibitem{dreyer2023large}
P.~Dreyer, A.~Roc, L.~Pillette, S.~Rimbert, and F.~Lotte.
\newblock A large eeg database with users’ profile information for motor imagery brain-computer interface research.
\newblock {\em Scientific Data}, 10(1):580, 2023.

\bibitem{GramfortEtAl2013a}
A.~Gramfort, M.~Luessi, E.~Larson, D.~A. Engemann, D.~Strohmeier, C.~Brodbeck, R.~Goj, M.~Jas, T.~Brooks, L.~Parkkonen, and M.~S. H{\"a}m{\"a}l{\"a}inen.
\newblock {{MEG}} and {{EEG}} data analysis with {{MNE}}-{{Python}}.
\newblock {\em Frontiers in Neuroscience}, 7(267):1--13, 2013.

\bibitem{jinpei2024}
J.~Han, X.~Gu, G.-Z. Yang, and B.~Lo.
\newblock Noise-factorized disentangled representation learning for generalizable motor imagery eeg classification.
\newblock {\em IEEE Journal of Biomedical and Health Informatics}, 28(2):765--776, 2024.

\bibitem{junqueira2024systematic}
B.~Junqueira, B.~Aristimunha, S.~Chevallier, and R.~Y. de~Camargo.
\newblock A systematic evaluation of euclidean alignment with deep learning for eeg decoding.
\newblock {\em Journal of Neural Engineering}, 21(3):036038, 2024.

\bibitem{kwon2020}
O.-Y. Kwon, M.-H. Lee, C.~Guan, and S.-W. Lee.
\newblock Subject-independent brain–computer interfaces based on deep convolutional neural networks.
\newblock {\em IEEE Transactions on Neural Networks and Learning Systems}, 31(10):3839--3852, 2020.

\bibitem{lachaux1999measuring}
J.-P. Lachaux, E.~Rodriguez, J.~Martinerie, and F.~J. Varela.
\newblock Measuring phase synchrony in brain signals.
\newblock {\em Human brain mapping}, 8(4):194--208, 1999.

\bibitem{lawhern2018eegnet}
V.~J. Lawhern, A.~J. Solon, N.~R. Waytowich, S.~M. Gordon, C.~P. Hung, and B.~J. Lance.
\newblock Eegnet: a compact convolutional neural network for eeg-based brain--computer interfaces.
\newblock {\em Journal of neural engineering}, 15(5):056013, 2018.

\bibitem{Ma2015}
B.~W. Ma~Minett.
\newblock Resting state eeg-based biometrics for individual identification using convolutional neural networks.
\newblock 2015.

\bibitem{mellot2023}
A.~Mellot, A.~Collas, P.~L.~C. Rodrigues, D.~Engemann, and A.~Gramfort.
\newblock {Harmonizing and aligning M/EEG datasets with covariance-based techniques to enhance predictive regression modeling}.
\newblock {\em Imaging Neuroscience}, 1:1--23, 12 2023.

\bibitem{rajpura2024}
P.~Rajpura, H.~Cecotti, and Y.~K. Meena.
\newblock Explainable artificial intelligence approaches for brain–computer interfaces: a review and design space.
\newblock {\em Journal of Neural Engineering}, 21(4):041003, aug 2024.

\bibitem{schalk2024translation}
G.~Schalk, P.~Brunner, B.~Z. Allison, S.~R. Soekadar, C.~Guan, T.~Denison, J.~Rickert, and K.~J. Miller.
\newblock Translation of neurotechnologies.
\newblock {\em Nature Reviews Bioengineering}, pages 1--16, 2024.

\bibitem{tallon1997oscillatory}
C.~Tallon-Baudry, O.~Bertrand, C.~Delpuech, and J.~Pernier.
\newblock Oscillatory $\gamma$-band (30--70 hz) activity induced by a visual search task in humans.
\newblock {\em Journal of Neuroscience}, 17(2):722--734, 1997.

\bibitem{trocellier2024validating}
D.~Trocellier, B.~N'Kaoua, and F.~Lotte.
\newblock Validating neurophysiological predictors of bci performance on a large open source dataset.
\newblock In {\em 9th Graz Brain-Computer Interface Conference 2024-GBCIC2024}, 2024.

\bibitem{shukahara2021translation}
T.~U. Tshukahara, Anzai.
\newblock A design of eegnet-based inference processor for pattern recognition of eeg using fpga.
\newblock {\em Electronics and Communications in Japan}, 1:53--64, 2021.

\bibitem{tzdaka2020}
E.~Tzdaka, C.~Benaroch, C.~Jeunet, and F.~Lotte.
\newblock Assessing the relevance of neurophysiological patterns to predict motor imagery-based bci users’ performance.
\newblock In {\em 2020 IEEE International Conference on Systems, Man, and Cybernetics (SMC)}, pages 2490--2495, 2020.

\bibitem{wang2019convolutional}
M.~Wang, H.~El-Fiqi, J.~Hu, and H.~A. Abbass.
\newblock Convolutional neural networks using dynamic functional connectivity for eeg-based person identification in diverse human states.
\newblock {\em IEEE Transactions on Information Forensics and Security}, 14(12):3259--3272, 2019.

\bibitem{zhang2021}
K.~Zhang, N.~Robinson, S.-W. Lee, and C.~Guan.
\newblock Adaptive transfer learning for eeg motor imagery classification with deep convolutional neural network.
\newblock {\em Neural Networks}, 136:1--10, 2021.

\bibitem{zhang2024torcheeg}
Z.~Zhang, S.~hua Zhong, and Y.~Liu.
\newblock {TorchEEGEMO}: A deep learning toolbox towards {EEG}-based emotion recognition.
\newblock {\em Expert Systems with Applications}, page 123550, 2024.

\end{thebibliography}

\end{document}